\documentclass[aps,a4paper,aps,pra,amsmath,amsfonts,superscriptaddress,twocolumn,10pt]{revtex4-1}
\usepackage{bm}                      
\usepackage{dcolumn}                 
\usepackage[table]{xcolor}
\usepackage{amsmath}
\usepackage{times}

\allowdisplaybreaks

\newcommand*{\centt}[1]{\multicolumn{1}{c}{#1}}

\newcolumntype{x}[1]{D{.}{.}{#1}}

\begin{document}
\title{Refractive index and generalized polarizability}

\author{Krzysztof Pachucki}
\affiliation{Faculty of Physics, University of Warsaw,
  Pasteura 5, 02-093 Warsaw, Poland}

\author{Mariusz Puchalski}
\affiliation{Faculty of Chemistry, Adam Mickiewicz University, Umultowska 89b, 61-614 Pozna{\'n}, Poland}

\begin{abstract}
  We investigate the role of retardation corrections to polarizability and to the refractive index.
  We found  that the classical electromagnetic theory of dielectrics requires corresponding modifications
  in terms of the nonlocality of the dielectric constant. This nonlocality should be taken into account
  in the interpretation of accurate measurements of the optical refractivity.
\end{abstract}

\maketitle

\section{Introduction}

A coherent monochromatic electromagnetic
plane wave propagates in a disordered medium of polarizable
particles, provided its wavelength is much larger than
the average distance between particles. Then the multiple scattering
has the only effect of changing the wavelength.
The ratio of this medium wavelength to the vacuum one is the refractive index. The
electromagnetic wave satisfies Maxwell's equations on the macroscopic level,
thus the refractive index is given by
the dielectric and magnetic constants of the medium.
If the wavelength is sufficiently short or a high-precision refractive index is intended,
the wave-vector  dependence of the  dielectric constant cannot be neglected.
High-precision measurements of the refractive index of a low-density helium gas
are being performed in several laboratories \cite{egan:17, jousten:17, silvestri:18}.

Helium is the best choice as it is the simplest noble-gas atom, thus its properties
can be calculated with the highest accuracy. The Clausius-Mosotti formula,
\begin{equation}
  \frac{\epsilon-\epsilon_0}{\epsilon+2\,\epsilon_0} = \frac{4\,\pi}{3\,\epsilon_0}\,\rho\,\alpha
\end{equation}
relates the dielectric constant to the microscopic property, namely, to the electric dipole
polarizability $\alpha$ of an atom. For this reason, the helium electric dipole polarizability
has been studied in a series of works summarized in Ref. \cite{puchalski:16},
resulting in a accuracy of about $0.1$ ppm,
which comes mainly from the estimation of unknown higher-order corrections.

In this work we demonstrate that the electric dipole polarizability is not sufficient
for a highly accurate description of the refractive index. On the fundamental level
it should be obtained from the forward photon-scattering amplitude. 
Apart from the electric and magnetic dipole polarizabilities, it includes also the so-called
retardation corrections in the form of a quadrupole and other types of polarizabilities.
This leads to the dependence of the refractive index on the wave-vector $\vec k$.
This dependence  affects the relation between the $\vec D$ and $\vec E$ fields,
which becomes nonlocal in the coordinate space [see Eq. (\ref{16})].
The wave-vector dependence of the dielectric constant
has already been indicated for the model systems of finite-size spherical insertions
in the electric dipole approximation \cite{ersfeld:98}.
Here, we derive a complete set of retardation corrections and present their accurate numerical values
for a gas of helium atoms. They affect the interpretation of the accurate measurements of the refractive index
\cite{egan:17, jousten:17, silvestri:18}.
As an additional result, we note that the dielectric tensor introduced in previous works
on this topic \cite{ersfeld:98,forcella:17} can be expressed in terms of scalar dielectric and magnetic
constants only.

\section{Photon propagator in the medium}
We will obtain only the leading term in the small $\rho$ expansion of the refractive index,
as it is sufficient for our purpose.
For this we consider the photon propagator in the presence of the noble gas atoms.
The photon wavelength $\lambda$ is assumed to be much larger than the atomic size $a_0$. Then,
the electromagnetic interaction can be expanded in powers of a small factor $a_0/\lambda$
and with inclusion of qubic terms takes the following form \cite{lwqed}
\begin{align}
\delta H =&\ \sum_a\Bigl\{-e\,\vec r_a\cdot\vec E
-\frac{e}{2}\,r^i_a\,r^j_a\,E^i_{,j}
-\frac{e}{6}\,r^i_a\,r^j_a\,r^k_a\,E^i_{,jk}
\nonumber \\ &\
+ \frac{e^2}{8\,m}\bigl(\vec r_a\times\vec B\bigr)^2
-\frac{e}{6\,m}\Bigl(L^i_a\,r^j_a+r^j_a\,L^i_a\Bigr)B^i_{,j}\Bigr\} \label{01}
\end{align}
where $E^i_{,j}$ denotes the derivative of $E^i$  with respect to $x^j$, and
the summation is over all the atomic electrons. The electromagnetic fields
in the above are taken at the position of the nucleus. Since the electron-nucleus mass ratio
is very small ($\sim 10^{-4}$ for He) we assume an infinite nuclear mass.
Equation (\ref{01}) demonstrates the existence of small nondipole couplings to the electromagnetic field
and will be used below in the derivation of the previously neglected retardation
correction to the refractive index of helium gas.  
We use theoretical units $\hbar=c=\epsilon_0 =1$ and $e^2= 4\,\pi\,\alpha_{\rm em}$ throughout this work.

Consider now the photon propagation through the medium of noble-gas atoms.
We assume, in accordance with planned measurements, that the photon frequency is much smaller than
the excitation energy of the helium atom. Then the scattering amplitude is real
and there is no photon absorption.
Each scattering on an individual atom is averaged over the whole space assuming a homogeneous
and isotropic density $\rho$ of atoms,
\begin{align}
  \sum_i e^{i\,(\vec k-\vec k')\,\vec R_i} =&\    \int  e^{i\,(\vec k-\vec k')\,\vec R}\,\rho\,d^3R
  \nonumber \\ =&\ \rho (2\,\pi)^3\,\delta^3(\vec k-\vec k'),
\end{align}
which shows that photon propagation depends only the elastic forward off-shell amplitude
for the scattering off the single atom, which is of the general form \cite{olmos_01}
\begin{equation}
  t^{ij} = 4\,\pi\,\big[\omega^2\,\alpha\,\delta^{ij} + k^2\,\chi\,(\delta^{ij}-\hat k^i\,\hat k^j)\big], \label{03}
\end{equation}
where $\alpha = \alpha(\omega, k)$ is the generalized electric polarizability,
and $\chi = \chi(\omega, k)$ is the magnetic one.
The above Eq. (\ref{03}) is the definition of these polarizabilities, and
in the nonrelativistic limit $\alpha$, $\chi$ coincides with the standard electric, magnetic dipole polarizability.
Moreover, we note that in the nuclear (particle) physics literature \cite{olmos_01},
the magnetic polarizability is denoted by $\beta$.

Let us now define $t'^{ij}=\rho\,t^{ij}$, $\alpha'=\rho\,4\,\pi\,\alpha$, and $\chi'=\rho\,4\,\pi\,\chi $
to be the corresponding densities. The free-photon propagator in the $A^0=0$ gauge is \cite{landau}
\begin{equation}
  G^{ij} = \frac{\delta^{ij}-\frac{k^i\,k^j}{\omega^2}}{\omega^2-k^2}
\end{equation}
Straightforward derivation of the photon propagator in the medium,
which does not exclude different atoms being at the same positions or multiple interaction with the same atom,
leads to 
\begin{align}
  \hat G_n =&\ \hat G - \hat G\,\hat t'\,\hat G + \hat G\,\hat t'\,\hat G\,\hat t'\,\hat G + \ldots \nonumber
  \\ =&\ \hat G\,(I+\hat t'\,\hat G)^{-1}
  \\ =&\
  \frac{I-\omega^{-2}\,k^2\,\hat k\otimes\hat k}
       {[\omega^2(1+\alpha')-k^2(1-\chi')]\,I-(\alpha'+\chi')\,k^2\,\hat k\otimes\hat k}, \nonumber        
\end{align}
This assumption that different atoms do not occupy the same position affects the second term
in the expansion of $n^2$ in powers of $\rho$ and  leads to the Clausius-Mossotti formula.
So our formalism is correct only up the the first term in $\rho$, the density of atoms.
An identity
\begin{equation}
  \bigl(A\,I + B\,\hat k\otimes\hat k\bigr)^{-1} =\frac{1}{A}\,
  \biggl(I - \frac{B}{A+B}\,\hat k\otimes\,\hat k\biggr)  
\end{equation}
with
\begin{align}
  A =&\ \omega^2(1+\alpha')-k^2(1-\chi'), \nonumber \\
  B =&\ -(\alpha'+\chi')\,k^2\,, \nonumber \\
  A+B=&\ (1+\alpha')\,(\omega^2-k^2).
\end{align}
gives
\begin{align}
  \hat G_n =&\ 
       \frac{I -\frac{(1-\chi')\,k^2}{(1+\alpha')\,\omega^2}\,\hat k\otimes\,\hat k}
            {[\omega^2(1+\alpha')-k^2(1-\chi')]}.            
\end{align}
We note that $\omega$ and $k$ are completely independent quantities here.

\section{Refractive index}
$G_n$ is the photon propagator in the medium and thus contains all the information
about the electromagnetic field. In particular, the relation between the photon frequency
and its wave-vector is obtained from the condition
\begin{equation}
  \omega^2(1+\alpha')-k^2(1-\chi') = 0,
\end{equation}
which corresponds to the pole in $\omega$ of the photon propagator at the fixed $k$.
The refractive index,  $\omega^2\,n^2 = k^2$ up to the terms linear in the density of atoms is
\begin{equation}
  n^2 = \frac{1+\alpha'}{1-\chi'}\approx 1+\alpha'+\chi' = 1+4\,\pi\,\rho\,(\alpha+\chi).
\end{equation}
Since polarizabilities $\alpha$ and $\chi$ depend on the frequency $\omega$ and the wave-vector $k$,
the above is a transcendental equation for $n$. However, in the leading order of the atomic density $\rho$,
polarizabilities for the propagation of light in the medium can be taken at $\omega = k$, namely,
$\alpha = \alpha(\omega, k) \approx\alpha(\omega,\omega)$
and $\chi = \chi(\omega, k) \approx\chi(\omega,\omega)$.
Accordingly, the forward-scattering amplitude for on-shell photons simplifies to
\begin{equation}
  t^{ij} = \omega^2\,4\,\pi\,[\alpha(\omega,\omega) + \chi(\omega,\omega)]\,\delta^{ij}\,. \label{12}
\end{equation}
We have not been able to derive the analog of the Clausius-Mossotti formula
using the above propagator formalism, so we limit ourselves only to terms which
are linear in the atomic density, although according to Ref. \cite{ersfeld:98}
such a generalization is possible.

For the general electromagnetic field the notion of the refractive index is less appealing,
because the wave equation no longer holds, as it is modified by the square of d'Alembertian. 
For example, the static Coulomb field is modified according to $4\,\pi/k^2/n(0,k)$ which is
different from the modification of the light propagation.

\section{Electrodynamics in matter}
The effective action for the electromagnetic field including medium in terms of a scattering amplitude
$t^{ij}$ from Eq. (\ref{03}) is
\begin{align}
  S =&\ \int\!d^4 x\,\frac{E^2-B^2}{2} +
  \int\!\frac{d^4k}{(2\,\pi)^4}\,A^{*i}(\omega,\vec k)\,t^{ij}(\omega,k)\,A^j(\omega,\vec k)\nonumber\\
  =&\ \int\!d^4 k \Bigl[\frac{\epsilon(\omega,k)}{2}\,\vec E^*\,\vec E
    -\frac{1}{2\,\mu(\omega,k)}\,\vec B^*\,\vec B\Bigr],
\end{align}
where
\begin{align}
  \epsilon(\omega,k) =&\ 1+4\,\pi\,\rho\,\alpha(\omega,k),\\
  \mu^{-1}(\omega,k) =&\  1-4\,\pi\,\rho\,\chi(\omega,k).
\end{align}
In general, the dependence of $\epsilon(\omega,k)$ on $k$ is to be interpreted
as a nonlocal relation between $\vec D$ and $\vec E$ fields \cite{forcella:17}, namely,
\begin{equation}
  \vec D(\omega,\vec r) = \int d^3r' \epsilon(\omega, \vec r-\vec r')\,\vec E(\omega,\vec r'), \label{16}
\end{equation}
similarly to the relation in the time domain. This nonlocal relation is obvious considering that
atoms have a finite size. Indeed, the $k$ dependence of $\alpha$ and $\chi$ goes with $a_0/\lambda$,
which is assumed to be a small factor.
However, when high precision is intended, as in Refs. \cite{egan:17, jousten:17, silvestri:18},
this nonlocal relation between $\vec D(\omega, \vec r)$ and $\vec E(\omega, \vec r)$ might play a role.
For this one needs to investigate the dependence of generalized atomic polarizabilities on $k$ and the consequences
on the Maxwell equations in the medium, for example, on boundary conditions close to metallic walls.
Here, we aim to estimate the magnitude of the dependence on $k$ and perform calculations
for the case of the helium atom.

\section{Retardation corrections to the generalized polarizability of the helium atom}
Assuming an infinite nuclear mass and a vanishing overall electron spin and overall angular momentum,
the interaction of the helium atom  with an electromagnetic field is given by Eq. (\ref{01}).
The scattering amplitude obtained from this interaction is given by three contributions,
 \begin{equation}
   t^{ij} = t_{\rm nrel}^{ij} + t_{\rm rel}^{ij} + t_{\rm ret}^{ij},
 \end{equation}
$t_{\rm nrel}$ is the scattering amplitude due to the nonrelativistic electric dipole polarizability \cite{landau},
 \begin{align}
t_{\rm nrel}^{ij} = \frac{2}{3}\,e^2\,\omega^2\,\alpha_0\,\delta^{ij},
 \end{align}
where $\alpha_0$ is defined below in Eq. (\ref{20}). 
 $t_{\rm rel}$ is due to the relativistic correction to the electric dipole polarizability and due to diamagnetic coupling,
 \begin{align} 
t_{\rm rel}^{ij} =&\
\frac{2}{3}\,e^2\,\omega^2\,\delta_{\rm rel}\alpha_0\,\delta^{ij}
+\frac{e^2}{6\,m}\,\sum_a\big\langle\vec r_a^{\,2}\big\rangle\,(k^2\,\delta^{ij}-k^i\,k^j),
 \end{align}
 where $\delta_{\rm rel}\alpha_0$ is a correction coming from the Breit-Pauli Hamiltonian
 (see Refs. \cite{pachucki:01,puchalski:16}).
 $t_{\rm ret}^{ij}$ is a retardation correction,
\begin{align}
t_{\rm ret}^{ij} =&\ e^2\,\omega^2\,\biggl[
  \frac{1}{18}\,\alpha_4\, k^i\,k^j
+\frac{1}{20}\,\alpha_1\,\Bigl(\delta^{ij}\,k^2+\frac{k^i\,k^j}{3}\Bigr)
\nonumber \\ &\
- \frac{2}{45}\,\alpha_2\,\bigl(\delta^{ij}\,k^2+2\,k^i\,k^j)
+\frac{1}{9}\,\alpha_3\,\bigl(\delta^{ij}\,k^2-\,k^i\,k^j)
\biggr],
\end{align}
where
\begin{align}
\alpha_0(\omega) \equiv&\ \frac{1}{2}\sum_{a,b}\bigg\langle r_a^k\biggl(\frac{1}{H-E+\omega}+\frac{1}{H-E-\omega}\biggr)r_b^k\bigg\rangle \label{20}\\
\alpha_1(\omega) \equiv&\ \frac{1}{2}\sum_{a,b}\bigg\langle (r_a^k\,r_a^l)^{(2)}\biggl(\frac{1}{H-E+\omega}
\nonumber \\ &\
                  +\frac{1}{H-E-\omega}\biggr)(r_b^k\,r_b^l)^{(2)}\bigg\rangle \\
\alpha_2(\omega) \equiv&\ \frac{1}{2}\sum_{a,b}\bigg\langle r_a^k
                     \biggl(\frac{1}{H-E+\omega}+\frac{1}{H-E-\omega}\biggr)r_b^k\,r_b^2\bigg\rangle \\
\alpha_3(\omega) \equiv&\ \sum_{a,b}\frac{1}{m}\,\bigg\langle r_a^k \frac{1}{(H-E+\omega)(H-E-\omega)}
\nonumber \\ &\
                   \times i\,(\vec L_b\times \vec r_b - \vec r_b\times\vec L_b)^k\bigg\rangle\\
\alpha_4(\omega) \equiv&\ \frac{1}{2}\sum_{a,b}\bigg\langle r_a^2\biggl(\frac{1}{H-E+\omega}+\frac{1}{H-E-\omega}\biggr)r_b^2\bigg\rangle
\end{align}
where $(r^i\,r^j)^{(2)} \equiv r^i\,r^j-\delta^{ij}\,r^2/3$ is the quadrupole moment operator.
Only the $t_{\rm ret}^{ij}$ is a new term, not considered yet in the context of the refractive index \cite{puchalski:16}.
The corresponding corrections to the generalized polarizabilities are
\begin{align}
\alpha_{\rm ret} =&\ \alpha_{\rm em}\,k^2\,\biggl(
   \frac{\alpha_1}{15} - \frac{2\,\alpha_2}{15} + \frac{\alpha_4}{18}\biggr) \\
\chi_{\rm ret} =&\ \alpha_{\rm em}\,\omega^2\,\biggl(
-\frac{\alpha_1}{60} + \frac{4\,\alpha_2}{45} + \frac{\alpha_3}{9} - \frac{\alpha_4}{18}\biggr)
\end{align}
where $\alpha_{\rm em}$ is the fine structure constant.
For small (real) photon frequencies and in atomic units
 \begin{align}
\alpha_{\rm nrel} =&\ \frac{2}{3}\,a_0^3\,\alpha_0(0),\\
 [\alpha+\chi]_{\rm ret} =&\ a_0^5\,k^2\,\biggl[
\frac{1}{20}\,\alpha_1(0) - \frac{2}{45}\,\alpha_2(0) +\frac{1}{9}\,\alpha_3(0)\biggr]
 \nonumber \\  \equiv &\  \alpha_{\rm nrel}\,(a_0\,k)^2\,\kappa,
 \end{align}  
 where the last equation is a definition of the dimensionless coefficient $\kappa$.

\section{Results and conclusions}
Numerical calculations for He (see Tab. I)  give $\kappa = -0.507$.
 With $a_0 = 0.529\cdot 10^{-10}$ m and $\lambda = 633$  nm \cite{egan:17, jousten:17},
 the retardation correction to the refractive index is equal to
 \begin{equation}
   \frac{\delta_{\rm ret} n}{n-1} = \Big(\frac{2\,\pi\,a_0}{\lambda}\Big)^2\,\kappa = -1.40\cdot 10^{-7}\,,
 \end{equation}
which is larger than the previously calculated $\omega^2$ term in the expansion of
the relativistic polarizability in the small $\omega$, $\delta n = -1.24\cdot 10^{-7}\,(n-1)$ \cite{puchalski:16}. 
We note that the retardation correction to the refractive index significantly depends on the
wavelength, and for the planned measurements at LNE \cite{silvestri:18} with $\lambda = 532$ nm
this correction amounts to $-1.98\cdot 10^{-7}$ which is larger than the anticipated accuracy of these measurements.
   
\begin{table}
\renewcommand{\arraystretch}{1.2}
\caption{Expectation values in a.u. in the helium atom}
\label{Toperators}
\begin{ruledtabular}
\begin{tabular}{l@{\extracolsep{\fill}}x{3.15}}
\centt{Polarizability} & \centt{Expectation value} \\ 
\hline\\[-1.5ex]               
$\alpha_0(0)$     &  2.074\,788\,3 \\
$\alpha_1(0)$     &  4.075\,138\,5 \\
$\alpha_2(0)$     &  9.317\,642\,6 \\
$\alpha_3(0)$     & -4.417\,669\,4 \\
$\kappa$          & -0.506\,952\,5
\end{tabular}        
\end{ruledtabular}   
\end{table}

\begin{acknowledgments}
The authors wish to thank Karol Makuch and Bogumi\l\ Jeziorski for interesting discussions.
This work was supported by the National Science Center (Poland) Grant No. 2017/27/B/ST2/02459.
\end{acknowledgments}

\end{document}